\documentclass[aps,pre,showpacs,twocolumn]{revtex4-2}
\usepackage{ulem}
\usepackage{xcolor}
\usepackage{graphicx}
\usepackage{dcolumn}
\usepackage{amsmath}
\usepackage{latexsym}
\usepackage{mathrsfs}
\usepackage{breqn}
\usepackage{tikz}
\usepackage{hyperref}
\hypersetup{
	colorlinks,
	citecolor=blue,
	filecolor=black,
	linkcolor=blue,
	urlcolor=black
}
\usepackage{mathtools}
\usepackage{amssymb}
\usepackage{cleveref}

\setcitestyle{super}

\newcommand{\be}{\begin{equation}} 
\newcommand{\ee}{\end{equation}} 
\newcommand{\bea}{\begin{eqnarray}} 
\newcommand{\eea}{\end{eqnarray}} 
\newcommand{\eps}{\epsilon}
\newcommand{\ra}{\rangle}
\newcommand{\la}{\langle}
\newcommand{\ts}{\textsuperscript}

\definecolor{darkgreen}{rgb}{0,.4,0}
\definecolor{mixedgreen}{rgb}{0.3,0.6,00}
\newcommand{\REMOVE}[1]%
           {{\color{magenta}\sout{#1}}}

\newcommand{\blackcircle}{\begin{tikzpicture}
                           \filldraw[fill=black,draw=black] circle (3pt);
                          \end{tikzpicture}
}
\newcommand{\whitecircle}{\begin{tikzpicture}
                           \filldraw[fill=white,draw=blue] circle (3pt);
                          \end{tikzpicture}
}

\begin{document}

\title{ {\Large \bf From Ising model to Kitaev Chain}\\
An introduction to topological phase transitions}
\author{Kartik Chhajed}  
\affiliation{Department of Physical Sciences IISER Mohali Knowledge city Sector 81 SAS Nagar Manauli 140306 India}

\begin{abstract}
 In this general article, we map the one-dimensional transverse field quantum Ising model of ferromagnetism to Kitaev's one-dimensional p-wave superconductor, which has its application in fault-tolerant topological quantum computing. Mapping Pauli's spin operators of transverse Ising chain to spinless fermionic creation and annihilation operators by Inverse Jordan-Wigner transformation leads to a Hamiltonian form closely related to Kitaev Chain, which exhibits topological phase transition where phases are characterized by different topological invariant that changes discontinuously at the transition point. Kitaev Chain supports two Majorana zero modes (MZMs) in the non-trivial topological phase, while none is in the trivial phase. The doubly degenerate ground state of the transverse Ising in ferromagnetic phase corresponds to non-local free fermion degree made from MZMs. The quasi-particle excitations of Ising chain, viz., domain wall formation in the ferromagnetic phase and spin-flip in paramagnetic phase maps to Bogoliubon excitations. The mapping suggests that a non-local order parameter can be defined for Kitaev Chain to work with the usual paradigm of Landau's theory.
\end{abstract}
\maketitle
    
\section{Introduction}
In 2001, Kitaev proposed a one-dimensional toy model containing a tight-binding chain of spinless electrons and a superconducting term\cite{majoranfermions}. Kitaev's one-dimensional p-wave superconductor, an example of a topological phase transition model, falls into two-dimensional Ising universality class for a given symmetry point where the model is exactly solvable\cite{PhysRevB.97.085131}. Kitaev Chain has its application in fault-tolerant topological quantum computing. One feature of Kitaev's open chain is that Majorana zero modes (MZMs) are present at its edges in one phase, while none is in another phase. These MZMs are topologically protected against decoherence by local perturbation and form the basic building block for topological quantum computing.

In the context of High energy physics, a fermion with property $a_j^\dagger = a_j$ implies that the particle's anti-particle is the particle itself. Mathematically, Majorana fermion is perfectly well defined. To this date, no Majorana fermions have been found existing as a fundamental particle. The nature of neutrinos is not settled, they may be either Dirac fermions or Majorana fermions. However, in condensed matter physics, they are found to exist as quasiparticles.

An anti-particle, in condensed matter physics, means devoid of an electron, i.e., a hole. In quantum computing, `electrons' and `holes' are encoded as q-bits:
$$\blackcircle \to |1\ra,\quad\whitecircle \to |0\ra$$
These q-bits are very sensitive to local perturbations. To remedy this caveat: two spatially separated Majorana bound states can be encoded as one fermionic degree of freedom in a very non-local way, making Majorana bound states topologically protected against decoherence by local perturbations. It is possible to experimentally design Kitaev's toy model using: a one-dimensional wire with appreciable spin-orbit coupling, conventional s-wave superconductor, and an external magnetic field \cite{NadjPerge}. 

In \cref{secquantIsing}, we start with the description of quantum cousin of the two-dimensional classical Ising model, viz., the transverse field quantum Ising chain, giving a brief overview of the phase transition phenomena occurring in the chain. Using Suzuki-Trotter formalism, it can be shown that the ground state of the $d$-dimensional transverse Ising model is equivalent to a specific $(d+1)$-dimensional classical Ising model \cite{10.1143/PTP.56.1454}. In \cref{mapquanttoclass},  as a special case, we have mapped the ground state of the one-dimensional transverse Ising model to the two-dimensional anisotropic Ising model by introducing imaginary time-slicing \cite{10.1143/PTP.46.1337}.

In \cref{sec:Isingtokitaev}, we mapped the transverse Ising chain to spinless fermionic theory via Inverse Jordan–Wigner Transformation \cite{PhysRev.76.1232, JordanWigner}. The Hamiltonian structure after transformation looks similar to famous Kitaev's one-dimensional p-wave superconductor Hamiltonian. However, the fermionic number is not conserved.  Since the Hamiltonian is quadratic (no interaction), it is possible to diagonalize Hamiltonian via Bogoliubov Transformation to a fermionic basis where the particle number would be conserved \cite{Valatin, Bogoljubov}. In \cref{phyofkitaev}, we study the physics of the Kitaev Chain, highlighting the emergence of topologically protected MZMs.
\section{Transverse field Quantum Ising model}\label{secquantIsing}
\begin{figure}[h]
 \caption{The coupling $g$ is a measure of the transverse field strength. For a large $g$ value, all spins point along the direction of the transverse field in the ground state. For a small $g$ value, the spins are ferromagnetically aligned by their exchange interactions.\label{figure2}}
 \centering
 \includegraphics[width=\columnwidth]{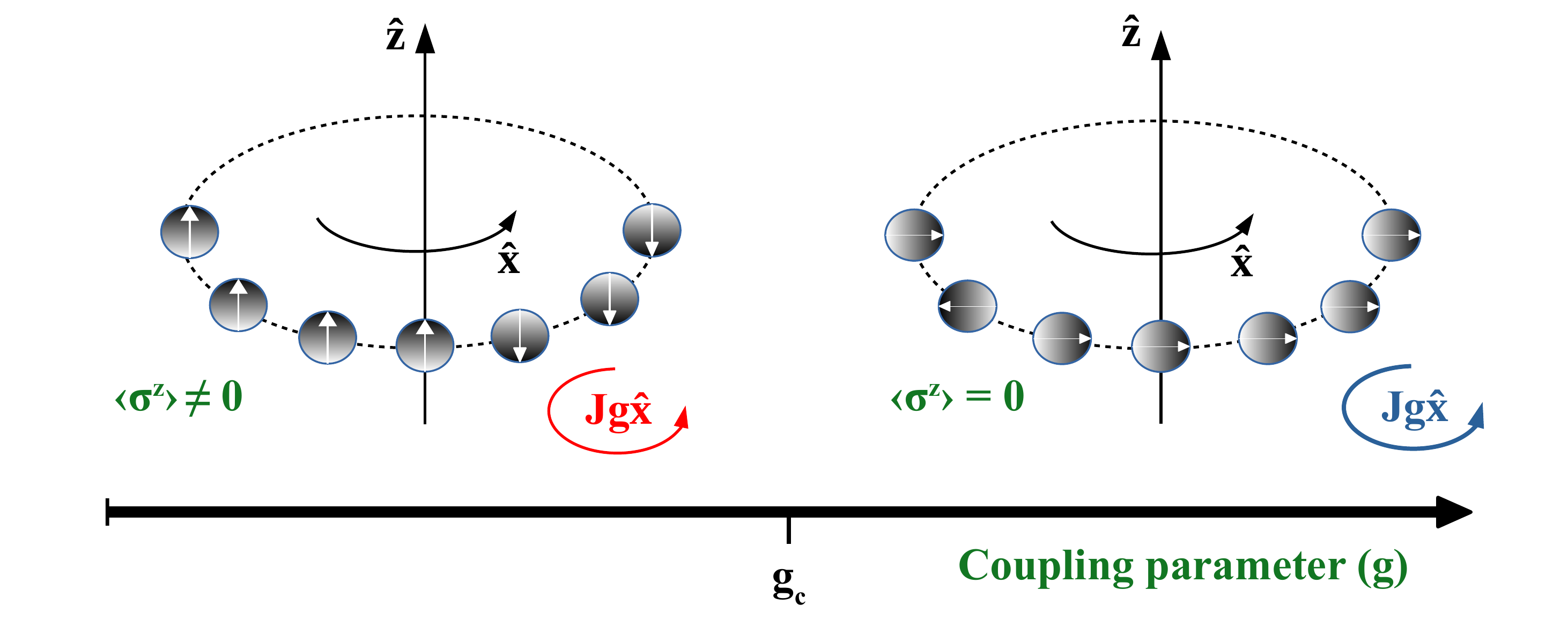}
\end{figure}

The transverse field quantum Ising model is considered as “Drosophila” of quantum phase transition. Unlike the classical Ising model, where thermal fluctuations drive phase transition, quantum fluctuations drive phase transition in the quantum Ising model. Consider the Hamiltonian for the one-dimensional transverse Ising model
\be\label{quantIsinghamil}
    {\cal H}_Q=-Jg\sum\limits_{i}\hat{\sigma}_i^x-J\sum\limits_{\la ij\ra}\hat{\sigma}_i^z\hat{\sigma}_j^z
\ee
where operators $\hat{\sigma}_i^{x,z}$ are the Pauli matrices at lattice site $i$. These operators commute at different sites, i.e., $[\sigma_i,\sigma_j]=0$ for $i\neq j$. The dimensionless $g$ is referred to as the coupling parameter. At zero temperature, the ferromagnetic interaction term $\hat{\sigma}_i^z\hat{\sigma}_{i+1}^z$, favors spin alignment ($\uparrow\uparrow$ or $\downarrow\downarrow$), whereas the field term $\hat{\sigma}_i^x$, favors spins to point in the positive $x$-direction ($\rightarrow$). The system switches between the disordered ($g \gg 1$) and the ordered ($g \ll 1$) phase as one tunes coupling parameter $g$, see \cref{figure2}. The point $g=g_c=1$ is the critical point where neither of the two descriptions is valid.

The Hamiltonian of transverse Ising chain is invariant under unitary transformation $\zeta=\prod\limits_i\sigma_i^x$, which flips all spins in the $z$-direction ($\mathbb{Z}_2$ symmetry)
$$\zeta|\uparrow\uparrow\downarrow\dots\ra = |\downarrow\downarrow\uparrow\dots\ra, \quad \zeta^2|\uparrow\uparrow\downarrow\dots\ra=|\uparrow\uparrow\downarrow\dots\ra$$
and since $\zeta^2=1$, and $\zeta\sigma^x_i\zeta=\sigma^x_i$ and $\zeta\sigma^z_i\zeta=-\sigma^z_i$, 
$$\therefore [{\cal H}_Q,\zeta] = 0$$
\subsection{Quantum Paramagnet}
\noindent When $g\gg 1$,
$${\cal H}_Q \backsimeq -Jg\sum\limits_{i}\hat{\sigma}_i^x$$
there is a unique ground state,
$$|\psi_0\ra = |\rightarrow\rightarrow\rightarrow\dots\ra;\quad\zeta|\psi_0\ra=|\psi_0\ra$$
Here the quasiparticle excitations which is $N$-fold degenerate, corresponds to a spin-flip in negative $x$-direction,
$$|\psi_{i}\ra=|\dots\rightarrow\rightarrow\underbrace{\leftarrow}_{\textrm{$i$\ts{th} spin-flip}}\rightarrow\rightarrow\dots\ra$$
The perturbation $V=-J\sum\hat{\sigma}_i^z\hat{\sigma}_{i+1}^z$ moves a spin-flip to its neighbouring sites:
$$\la \psi_i |V|\psi_j\ra = -J(\delta_{j,i-1}+\delta_{j,i+1})$$
The effective Hamiltonian in a single spin-flip basis is
$${\cal H}_{eff}|\psi_i\ra = -J(|\psi_{i-1}\ra+|\psi_{i+1}\ra) + (E_0+2gJ)|\psi_{i}\ra$$
For diagonalizing ${\cal H}_{eff}$, we do the Fourier transformation
$$|\psi_{k}\ra = \frac{1}{\sqrt{N}}\sum\limits_je^{-ikj}|\psi_{j}\ra$$
Then,
$$({\cal H}_{eff}-E_0)|\psi_{k}\ra = 2J(g-\cos k)|\psi_{k}\ra= \epsilon_k |\psi_{k}$$
In the long-wavelength limit, the quasiparticle excitation energy is $\epsilon_k \backsimeq \Delta + Jk^2 $, where $\Delta=2J(g-1)$ is the bulk energy gap, see \cref{gapquantumIsing}.

\subsection{Quantum ferromagnet}\label{Quantumferromagnet}
\noindent In limit $g\to 0$,
$${\cal H}_Q \backsimeq -J\sum\limits_{\la ij\ra}\hat{\sigma}_i^z\hat{\sigma}_j^z,$$
there are two degenerate ground states
$$|\psi_\uparrow\ra=|\uparrow\uparrow\uparrow\dots\ra,\textrm{ and }|\psi_\downarrow\ra=|\downarrow\downarrow\downarrow\dots\ra$$
These ground states does not respect the symmetry
$$\zeta|\psi_\uparrow\ra=|\psi_\downarrow\ra\textrm{ and }\zeta|\psi_\downarrow\ra=|\psi_\uparrow\ra$$
However, the linear combination of macroscopic ket states, 
$$|\psi_{\pm}\ra=\frac{|\psi_\uparrow\ra\pm|\psi_\downarrow\ra}{\sqrt{2}};\quad \zeta|\psi_{\pm}\ra=\pm |\psi_{\pm}\ra$$
preserve the symmetry. The degeneracy is lifted by $N$\textsuperscript{th}-order perturbation theory with perturbation $V = -Jg\sum\hat{\sigma}_i^x$, and $N$ is number of Ising spins in the chain. The effective Hamiltonian in $|\psi_{\uparrow\downarrow}\ra$ basis is
$${\cal H}_{eff} = \begin{pmatrix}
  E_0& g^N \\
  g^N& E_0
  \end{pmatrix}
$$
The true ground state is $|\psi_+\ra$ with exponentially small splitting of $\delta=e^{-N\ln(1/g)}$ with $|\psi_-\ra$. At time $t=0$, if we prepare system in state $|\psi(t=0)\ra=|\psi_\downarrow\ra$, after time $t$ the system will be in state 
$$|\psi(t)\ra = \frac{e^{-\dot\iota E_+t}|\psi_+\ra-e^{-\dot\iota E_-t}|\psi_-\ra}{\sqrt{2}}$$
The probability of finding system in state $|\psi_\downarrow\ra$ will be
$$P(t) = |\la \psi_{\downarrow} |\psi(t)\ra|^2 = \cos^2\bigg(\frac{\delta t}{2}\bigg)\backsimeq 1 \textrm{ for $t\ll 1/\delta \approx e^{N\ln(1/g)}$}$$
So, for $N=N_{\textrm{Avogadro}}$, the initial state $|\psi_\downarrow\ra$ is the true ground state, unless one is willing to wait for time, $t=e^{10^{23}}$ to see tunneling process to state $|\psi_\uparrow\ra$. The quasiparticle excitations in ferromagnetic phase are the formation of domain-walls
$$\phi_{\bar i=i+\frac12}=|\dots\uparrow\underbrace{\uparrow}_i\vdots\underbrace{\downarrow}_{i+1}\downarrow\downarrow\dots\dots\ra$$
The domain-wall formation comes in pairs (at the independent sites) to respect the periodic boundary condition. In this phase, the perturbation $V=-J\sum\hat{\sigma}_i^x$ moves a domain-wall to its neighbouring sites. If we follow calculation similar to the paramagnetic phase, we will get the quasiparticle excitation energy as $\eps_k = 2J(1-g\cos k) \backsimeq \Delta + Jk^2 $ in ferromagnetic phase. Here the bulk energy gap for domain-wall formations is $\Delta=2J(1-g)$.

\begin{figure}[!h]
\caption{Plot of energy dispersion for different values of $g$. The gap $\Delta$ closes at $g=1$}
\label{gapquantumIsing}
\centering
\includegraphics[width=\columnwidth]{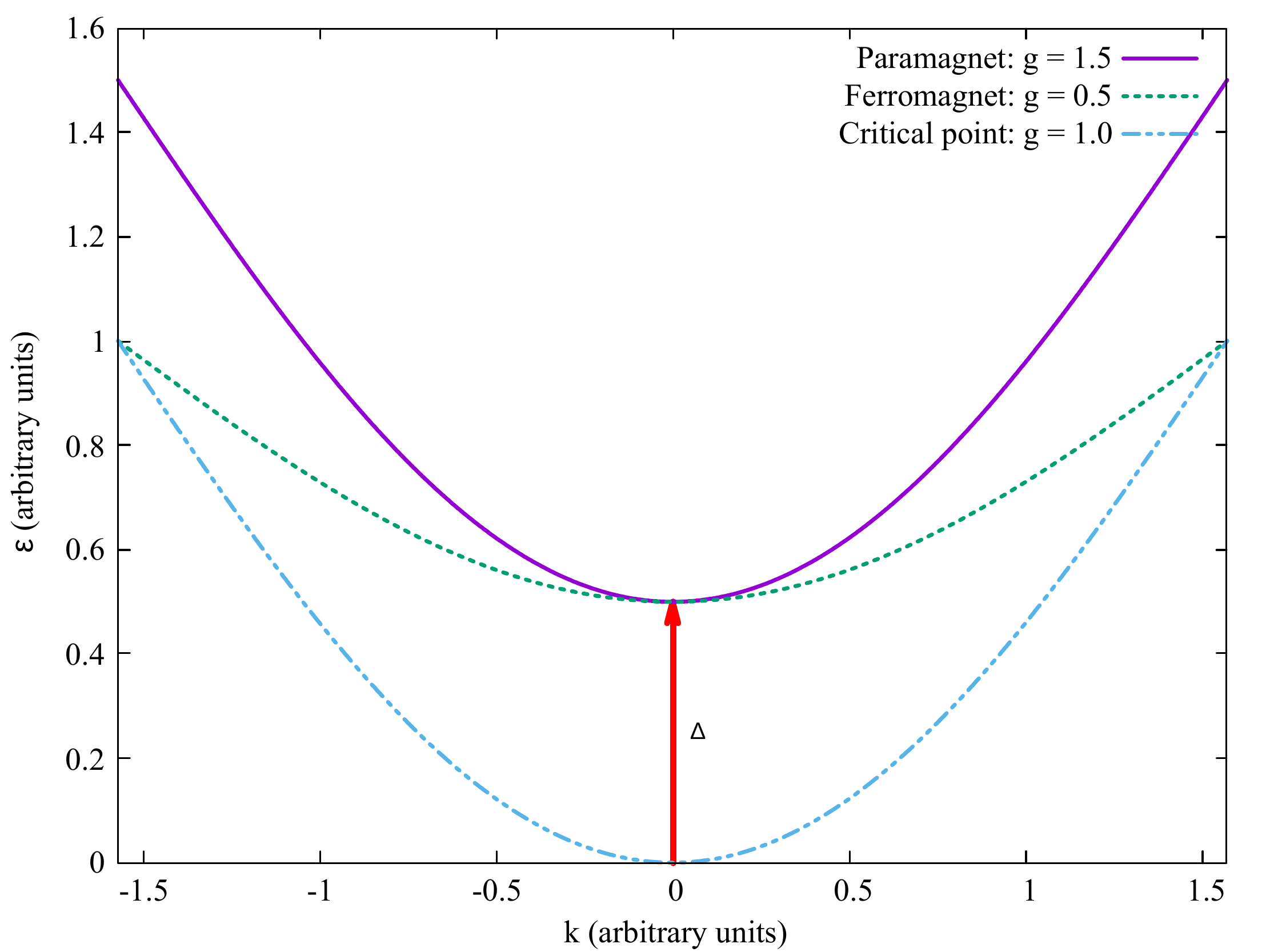}
\end{figure}

In \cref{gapquantumIsing}, the bulk gap $\Delta=2J|1-g|$, closes at $g=1$. For strongly coupled transverse Ising and weakly coupled transverse Ising, the symmetric gap function $\Delta=2J|1-g|$ suggests duality between two phases, shown more formally as Kramers–Wannier duality.

\subsection{Kramers–Wannier duality}
\noindent We define domain-wall variables $\varsigma_{\bar i}^x$ and $\varsigma_{\bar i}^z$ as
$$\varsigma_{\bar i}^x := \sigma_{i}^z\sigma_{i+1}^z \textrm{ and } \varsigma_{\bar i}^z:=\prod\limits_{i>\bar i}\sigma_i^x ;\quad \bar i = i + \frac12$$
Where,
$$\varsigma_{\bar i}^x =\begin{cases}-1, & \textrm{if domain-wall at $\bar i$}\\
                                     +1, & \textrm{otherwise}\end{cases}$$
and $\varsigma_{\bar i}^z$ create a domain-wall by flipping all spin to the right of $\bar i$. The domain-wall variables follow the Pauli matrix algebra. A combination of $\sigma$-spin variable and $\varsigma$-spin variable is non-local. E.g., Majorana variables \cite{PhysRev.76.1232}, defined as
$$a_i=\sigma_i^z\varsigma_{\bar i}^z,\quad b_i=\sigma_i^y\varsigma_{\bar i}^z$$
Majorana variables are non-local because of string $\Pi\sigma^x$. We will come back to this in \cref{phyofkitaev}. On writing the quantum Ising Hamiltonian, see \cref{quantIsinghamil}, in terms of domain-wall variables, we get
$${\cal H}_Q=-J\sum\limits_{\bar i}\hat{\varsigma}_{\bar i}^x-Jg\sum\limits_{\la {\bar i}{\bar j}\ra}\hat{\varsigma}_{\bar i}^z\hat{\varsigma}_{\bar j}^z$$
The same Hamiltonian but with different coupling
$$J\leftrightarrow gJ$$
The paramagnetic phase of $\varsigma$-spins corresponds to the ferromagnetic phase of $\sigma$-spins, and vice-versa. The point $g=1$ is the self-dual point. Albeit, the ground state of the paramagnetic phase of $\varsigma$-spins is unique, while the ground state of the ferromagnetic phase of $\sigma$-spins is doubly degenerate. It is because $\sigma$-spin description to the domain-wall description is two-to-one mapping. For e.g., in $\sigma$-spin description, states $|\uparrow\uparrow\vdots\downarrow\downarrow\dots\ra$ and $|\downarrow\downarrow\vdots\uparrow\uparrow\dots\ra$ maps to single domain-wall variable.

\section{Getting the Kitaev Chain from Ising model}\label{sec:Isingtokitaev}
In the following section, we will perform the Inverse Jordan-Wigner transformation to rewrite the spin variables as fermionic variables. Using Inverse Jordan-Wigner transformation, we will map the transverse Ising system to a system of spinless fermions. It involves rewriting the Pauli matrices so that they look like creation and annihilation operators. Nevertheless, it has a caveat. The number operator does not commute with the Hamiltonian. Since the Hamiltonian is quadratic (no interactions), we can diagonalize Hamiltonian by doing Bogoliubov transformation, which will take care of the problem, and we can study bulk property.
\subsection{The Jordan-Wigner transformation}\label{JWTranssection}
\noindent We define raising and lowering operators for Ising chain
$$\hat{\sigma}^{\pm}_i=\frac12(\hat{\sigma}_i^x\pm\dot\iota\hat{\sigma}_i^y)$$
which satisfy the anti-commutation relations
$$\{\sigma^-_i,\sigma^+_i\}=1,\quad \{\sigma^-_i,\sigma^-_i\}=\{\sigma^+_i,\sigma^+_i\}=0$$
The raising and lowering operators flip the spin
$$\hat\sigma^-|\uparrow\ra=|\downarrow\ra, \qquad\hat\sigma^+|\downarrow\ra=|\uparrow\ra$$
Suppose we define a spin-up state as a hole and a spin-down state as a particle, viz., 
$$|\uparrow\ra\equiv\whitecircle, \quad |\downarrow\ra\equiv\blackcircle$$
In that case, we are tempted to identify raising and lowering operators as creation and annihilation operators:
$$\hat\sigma^-\to c^\dagger,\qquad\hat\sigma^+\to c$$

\begin{figure}[h]
\caption{Realization of spins of quantum Ising model as hard-core bosons on a chain.}\label{Isingbosons}
\centering
\includegraphics[width=\columnwidth]{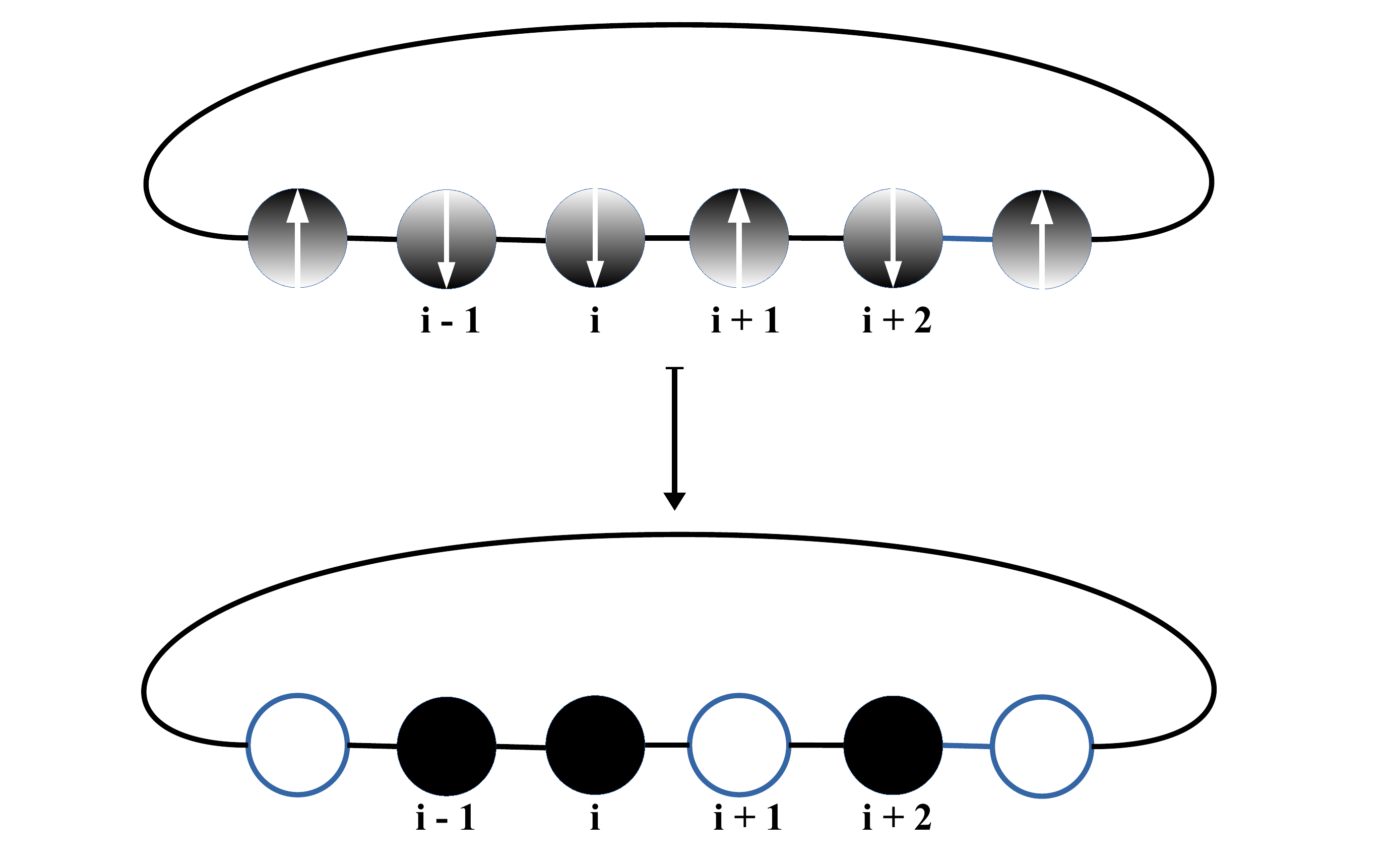}
\end{figure}

The spin operators $\hat{\sigma}^{\pm}_i$ and $\hat{\sigma}^{z}_i$ are generators of lie algebra isomorphic to creation and annihilation operators $c_i^\dagger,c_i$ and $n_i\equiv c_i^\dagger c_i$. Since, a spin can be flipped only once
$$\hat\sigma^-(\hat\sigma^-|\uparrow\ra)=0, \qquad\hat\sigma^+(\hat\sigma^+|\downarrow\ra)=0$$
A spin can be realized either as a hard-core boson or a fermion satisfying Pauli's exclusion. The fermionic operators are anticommutative at different sites, i.e., $\{c_i,c_j\}=\{c_i^\dagger,c_j^\dagger\}=0$, whereas the raising and lowering operators commute at different sites (bosonic)
$$[\sigma^+_i,\sigma^-_j]=0,\quad [\sigma^-_i,\sigma^+_j]=0;\qquad i\neq j$$
Therefore, the raising and lowering operators should be treated appropriately as the creation and annihilation operators of hard-core bosons, see \cref{Isingbosons}.

In 1928, Jordan and Wigner \cite{JordanWigner} performed a transformation which recovers the true fermions anticommutation relations from spin-operators
$$c_i=\prod\limits_{j<i}(\hat{\sigma}_j^z)\hat{\sigma}_i^+,\quad c_i^\dagger=\prod\limits_{j<i}(\hat{\sigma}_j^z)\hat{\sigma}_i^-$$
Note that, the fermionic operators are non-local since they depend on the state on each lattice site. The spin operator is only defined at a point (local), while the fermionic operator depends on the spin values along a whole line starting from the left boundary and ending at a given location. The Jordan–Wigner transformation can be inverted by identifying $\hat{\sigma}_i^z=1-2c_i^\dagger c_i$
$$\hat{\sigma}_i^+=\prod\limits_{j<i}(1-2c_j^\dagger c_j)c_i,\quad \hat{\sigma}_i^-=\prod\limits_{j<i}(1-2c_j^\dagger c_j)c_i^\dagger$$
It is convenient to rotate the spin axes by an angle $\pi/2$ about $y$-axis to further simplify the algebra, so that, $\hat{\sigma}_i^z\to\hat{\sigma}_i^x$ and $\hat{\sigma}_i^x\to -\hat{\sigma}_i^z$. In this frame, the Pauli matrices $\hat{\sigma}_i^z=-(\hat{\sigma}_i^++\hat{\sigma}_i^-)$ and the Pauli matrices $\hat{\sigma}_i^{x,z}$ in terms of fermionic operators is

$$\hat{\sigma}_i^x=1-2c_i^\dagger c_i,\quad \hat{\sigma}_i^z=-\prod\limits_{j<i}(1-2c_j^\dagger c_j)(c_i+c_i^\dagger)$$
Substituting expression for $\sigma_i^{x,z}$ in the Hamiltonian of quantum Ising model in \cref{quantIsinghamil}, we get
\begin{multline}
\label{eq:afterJWT}
    {\cal H}_{\textrm{JW}}=-J\sum\limits_i\Big(c_{i}^\dagger c_{i+1} + c_{i+1}^\dagger c_{i} - 2gc_{i}^\dagger c_{i}+g\\
    + c_{i}^\dagger c_{i+1}^\dagger + c_{i+1} c_{i}\Big)
\end{multline}
As mentioned earlier, because of the terms like $c_{i}^\dagger c_{i+1}^\dagger$ and $c_{i+1}c_{i}$, the fermionic number is not conserved. Nevertheless, since the additional terms are quadratic in the fermionic operator, we can diagonalize the Hamiltonian. However, we should be careful about the boundary condition.  If the spin chain has a periodic boundary condition, then the fermionic chain has an anti-periodic (periodic) boundary condition if there are even (odd) number of fermions \cite{boundarycondition}. The open boundary Ising model maps to an open boundary fermionic chain.

Assuming the system is large, one expects the chain’s interior to be the same for both boundary conditions. The key difference is the appearance of MZMs on the two ends in the open chain. The closed chain has a unique ground state, while the open-chain has a degenerate ground state.

The Hamiltonian in \cref{eq:afterJWT} is very similar to the famous Kitaev's one-dimensional p-wave superconductor Hamiltonian:
\begin{multline*}
{\cal H}_{\textrm{Kitaev}}=\sum\limits_{j=1}^{N}\Bigg[\underbrace{-\frac{t}{2}(c_{j+1}^\dagger c_j+c_{j}^\dagger c_{j+1})}_{\textrm{tight binding}}-\underbrace{\mu c_{j}^\dagger c_j}_{\textrm{chemical potential}}\\
+\underbrace{\frac{\Delta}{2}(c_{j}^\dagger c_{j+1}^\dagger + c_{j+1} c_{j})}_{\textrm{mean field p-wave superconducting term}}\Bigg]
\end{multline*}
For $\mu > t$, system forms a non-topological phase without Majorana modes in the open chain. Whereas for $\mu<t$, a topological phase emerges with MZMs in the open chain. For $\Delta,t>0$, the model belongs to the Ising universality class. In particular case $t=\Delta$, we identify Kitaev’s Hamiltonian same as quantum Ising Hamiltonian. For $\Delta = 0, t>0$, the model reduces to XX chain, viz., the isotropic limit of the XY model \cite{PhysRevB.97.085131, Franchini2017AnIT}.
\subsection{The Bogoliubov Transformation}\label{Bogoliubovtrans}
In 1958, Nikolay Bogoliubov and John George Valatin independently developed the Bogoliubov transformation for finding solutions of BCS theory in a homogeneous system \cite{Valatin, Bogoljubov}. Before the Bogoliubov transformation, first, we perform lattice Fourier transform with
$$c_k=\frac{1}{\sqrt{N}}\sum\limits_j c_je^{ikx_j}$$
where $x_j=aj$ and $a$ is the lattice constant set to unity for simplicity. After transformation, the \cref{eq:afterJWT} becomes
\begin{multline}
\label{afterfourier}
{\cal H}_f=\sum\limits_k\bigg(2[Jg-J\cos(k)]c_{k}^\dagger c_{k}\\
+iJ\sin(k)[c_{-k}^\dagger c_{k}^\dagger + c_{-k}c_{k}]-Jg\bigg)
\end{multline}
Ignoring the constant term, the Hamiltonian in \cref{afterfourier} can also be written in the standard Bogoliubov-de Gennes form
\be\label{bdgequation}
{\cal H}_{BdG} = J\sum\limits_k\Psi_k^\dagger \begin{pmatrix}g-\cos k & -i \sin k\\ i\sin k & -g+\cos k \end{pmatrix}\Psi_k
\ee
where
$$\Psi_k = \begin{pmatrix}
           c_{-k}\\
           c_k^\dagger
           \end{pmatrix}
$$
The particle-hole operator $\cal P$, exchanges the creation and annihilation parts of $\Psi_k$,
$${\cal P}=\prod\limits_k\tau^x_k\kappa;\quad \tau^x=\begin{pmatrix}
                0&1\\
                1&0
               \end{pmatrix}$$
where $\kappa$ is the complex conjugation operator. The squared particle-hole operator is ${\cal P}^2=+1$ and the system show particle-hole symmetry, $\{{\cal H}_{BdG},{\cal P}\}=0$. Given a solution with energy $\eps$ and momentum $k$, particle-hole symmetry  dictates, in general, the presence of a solution with energy $-\eps$ and momentum $-k$.

Now, we will diagonalize the Hamiltonian by doing the Bogoliubov transformation. We define Bogoliubons $\gamma_k$ as
$$\gamma_k:=u_kc_k-iv_kc_{-k}^\dagger$$
A Bogoliubon is a mixture of a electron and a hole, satisfying fermionic creation and annihilation anti-commutation relations 
$$\{\gamma_k,\gamma_l^\dagger\}=\delta_{kl};\, \{\gamma_k^\dagger ,\gamma_l^\dagger \}=\{\gamma_k,\gamma_l\}=0$$
where $u_k$ and $v_k$ satisfies the following property
$$u_k^2+v_k^2=1;\, u_{-k}=u_k;\, v_{-k}=-v_k$$
The following choice of $u_k$ and $v_k$ suffices the property:
$$u_k=\cos\bigg(\frac{\theta_k}{2}\bigg),\,v_k=\sin\bigg(\frac{\theta_k}{2}\bigg),\tan(\theta_k)=\frac{\sin(k)}{g-\cos(k)}$$
On substituting $c_k$ in \cref{afterfourier}, we get diagonalized Hamiltonian
\be\label{eqBVtrans}
 {\cal H}=\sum\limits_k\eps_k(\gamma_k^\dagger \gamma_k-\frac12);\ \eps_k=2J\sqrt{1+g^2-2g\cos(k)}
\ee

\begin{figure}[h]
\centering
\caption{Plot of energy dispersion $\eps_k$ against crystal momentum $k$ for different values of coupling parameter $g$.}\label{fig:epislonk}
\includegraphics[width=\columnwidth]{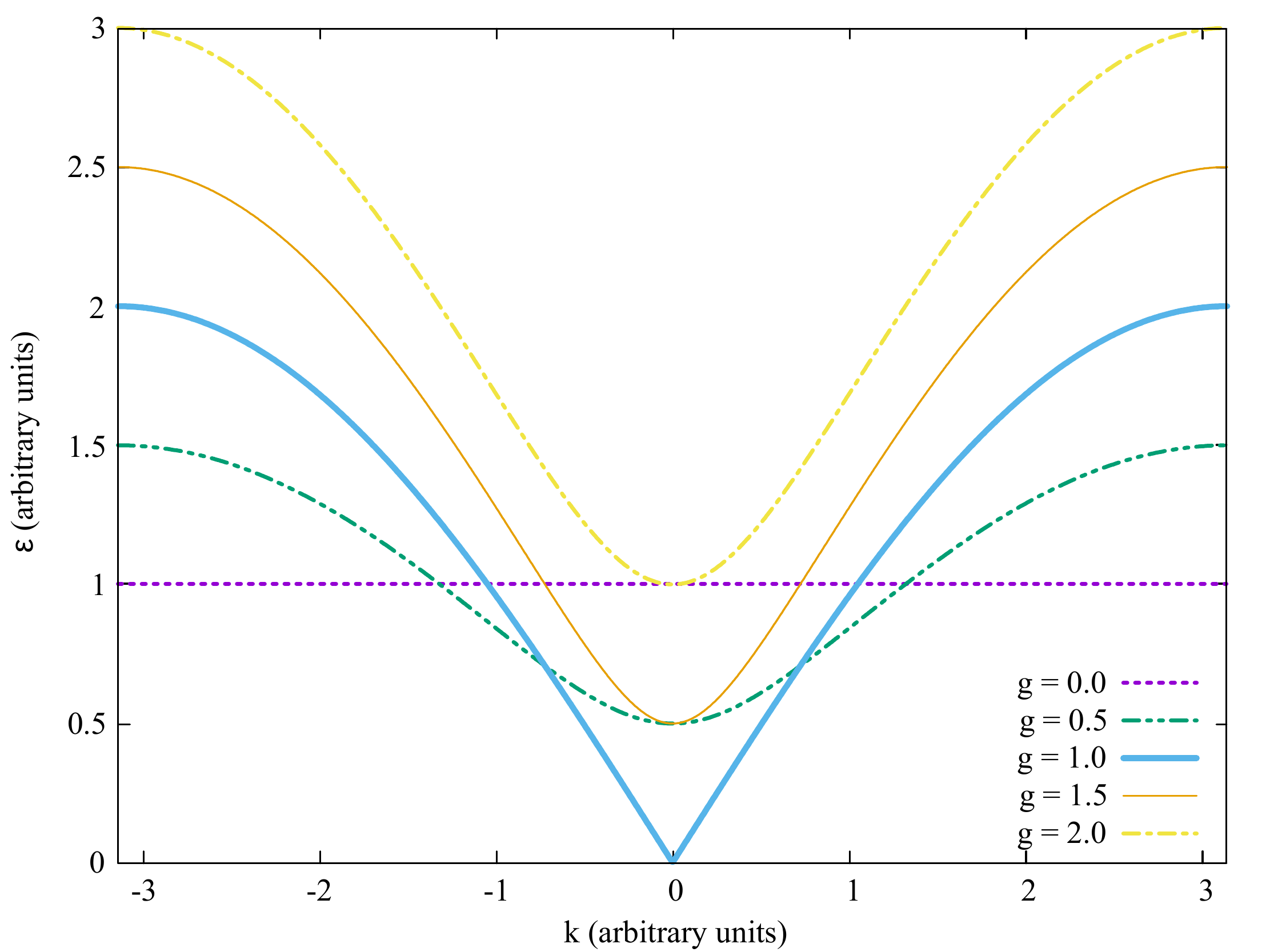}
\end{figure}

The energy gap vanishes at criticality, $g=g_c=1$, see \cref{fig:epislonk}. In the long-wavelength limit,
$$\eps_{k}=2J\sqrt{(1-g)^2+(k)^2}$$
At critical point $g=g_c$, the energy gap goes as 
$$\eps_k = 2J|k|$$ 
We get dynamical critical exponent $\eps_k\sim k^z,\,z=1$. We see emerging Dirac physics where bulk gap closes. In the long-wavelength limit, the Bogoliubov-de Gennes Hamiltonian in \cref{bdgequation} can be written in terms of Pauli matrices
$${\cal H}_{BdG}=\sum\limits_k\Psi_k^\dagger{H}_{Dirac}(k)\Psi_k ;\quad{H}_{Dirac}(k) = m\tau^z + Jk\tau^y$$
with mass in natural units, $m = J(g-1)$. When $m\to 0$, there are two energy eigenstates with energy $E=\pm Jk$, which are also eigenstates of $\tau^y$. That means eigenstates are the equal superposition of electrons and holes. They are Majorana modes free to propagate in the chain with speed $v = J $. In the massless limit, the free fermionic theory is conformally invariant, and one can use conformal field theory and operator product expansion to extract other critical exponents \cite{Ginsparg1988APPLIEDCF}.

\section{Physics of Kitaev Chain}\label{phyofkitaev}
\noindent One can formally define Majorana operators as
$$a_{j}=c_j^\dagger+c_j,\quad b_{j}=\dot\iota(c_j-c_j^\dagger)$$
with the properties
$$a_j^\dagger = a_j, \quad b_j^\dagger = b_j$$
$$\{a_j,a_{j'}\}=2\delta_{jj'};\quad \{b_j,b_{j'}\}=2\delta_{jj'};\quad \{a_j,b_{j'}\}=0$$
It is instructive to think that two Majorana modes describe one fermion. On rewriting the Hamiltonian in \cref{eq:afterJWT} in terms of the Majorana operators, we get
\be\label{Majoranaeq}
{\cal H} = \dot\iota J\sum\limits_{j}(a_{j}b_{j+1}+ga_{j}b_{j}) 
\ee

\begin{figure}[h]
\caption{Majorana Chain in two limits: $g \gg 1$ and $g\to 0$. In the former limit, the Majoranas pair-up at the same lattice site. In the latter limit, Majoranas pair-up at adjacent lattice sites leaving two `unpaired' Majorana zero-modes $b_0$ and $a_N$ at the ends of the chain.}\label{majoranachain}
\centering
\includegraphics[width=\columnwidth]{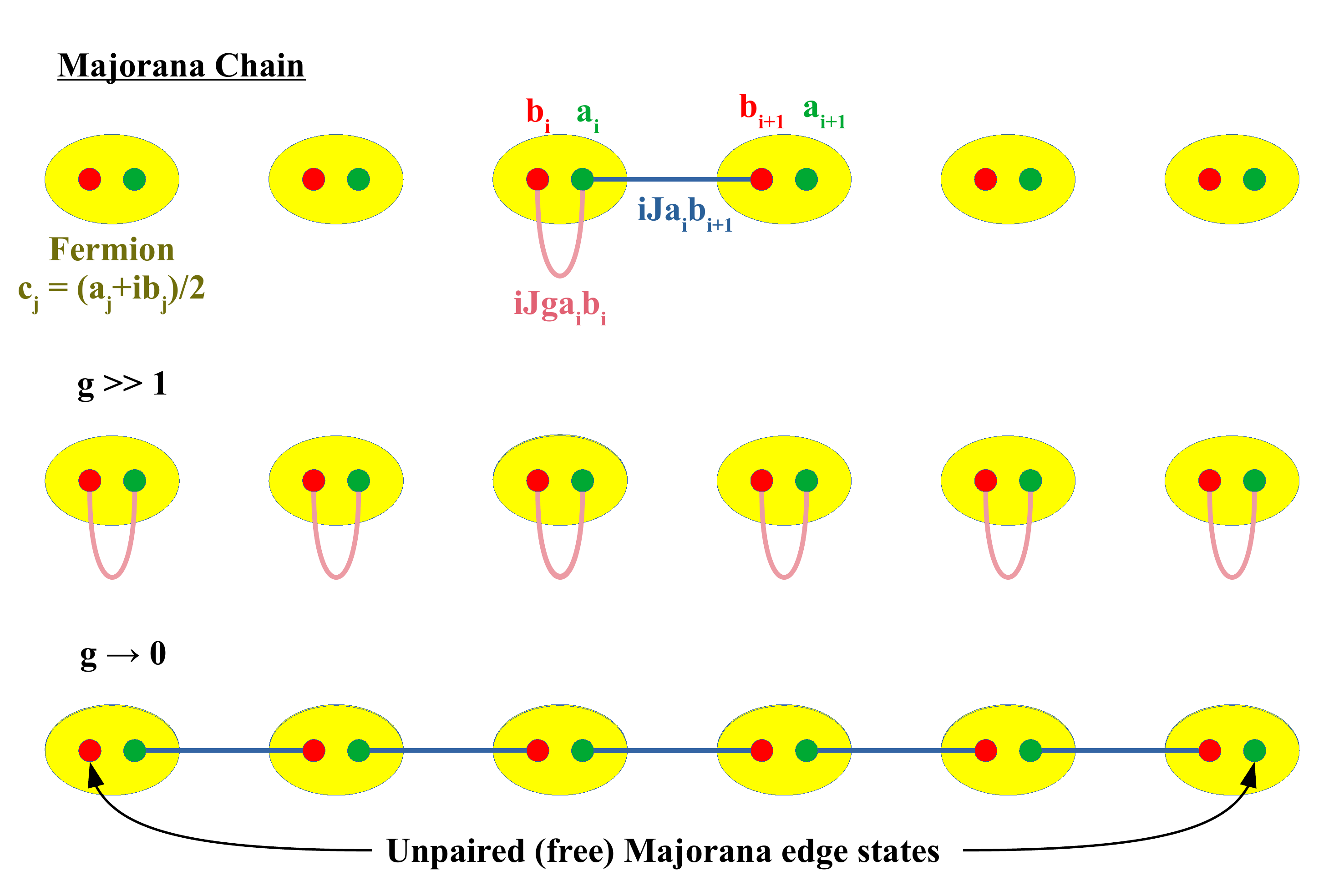}
\end{figure}

\begin{figure}[h]
\caption{The plot of dispersion relation for particle-hole symmetric Hamiltonian for the different value of coupling parameter $g$. The bulk gap closes at $g=1$.}\label{topologyandgap}
\centering
\includegraphics[width=\columnwidth]{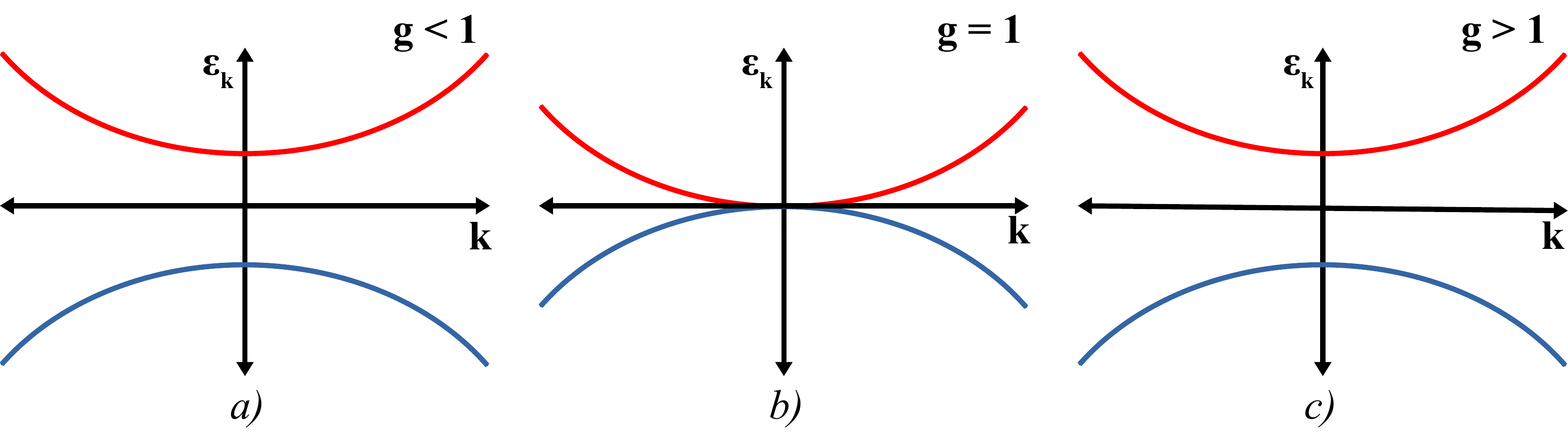}
\end{figure}

In the limit $g\gg 1$, the coupling dominates between Majorana modes $a_i$ and $b_i$ at the same lattice site with no zero-energy edge states (trivial), see \cref{majoranachain}. The energy cost for each Majorana pair is $gJ$, and the chain has a gaped bulk, see \cref{topologyandgap}.

In the limit $g\to 0$, the coupling dominates between Majorana modes at the adjacent lattice sites, and the two ends of chain support `unpaired' zero-energy Majorana modes (non-trivial), see \cref{majoranachain}. The energy cost for each paired Majorana modes is $J$, and the chain has a gaped bulk, see \cref{topologyandgap}. 

However, it should be noted that the MZMs solutions are not restricted to the extreme limit $g\to 0$. The particle-hole energy spectrum is symmetric around zero energy. In the limit $g\to 0$, we have two zero energy levels, corresponding to the MZMs, which are localized far away from each other and separated by a gaped medium (bulk). It is not possible to move these levels from zero energy individually (as one needs to respect particle-hole symmetry). The only way to split the Majorana modes in energy is by closing the bulk energy gap\cite{Reso3}.

In mathematics, two objects are considered topologically inequivalent if they cannot be continuously deformed into one another without tearing apart or gluing together parts. E.g., it is not possible to mold a frisbee into a donut without hollowing it. Analogous to that, it is not possible to go from the phase that supports MZMs to the phase that does not (or vice-versa) without closing the bulk gap. As long as the bulk gap is open, the MZMs are protected. This kind of topological protection is a generic characterization of topological edge modes that define the topological phase. Entering and exiting the topological phase requires closing the bulk gap, referred to as topological quantum phase transition.

In \cref{majoranachain}, for topological non-trivial phase, we get two free Majorana edge modes for free boundary condition, i.e., $[H,b_0] = [H,a_N] = 0$. We can construct a non-local fermionic mode from edge states
$$d^\dagger = \frac{b_0+ia_N}{2}$$
The $d$-fermion can either be occupied or empty, corresponding to two degenerate ground states. The doubly degenerate ground state of the ferromagnetic phase of the transverse Ising model maps onto a topologically non-trivial phase Kitaev Chain with doubly degenerate ground states. The $d$-fermion can be used as q-bits
$$|\uparrow\ra = d^\dagger|0\ra;\qquad |\downarrow\ra = |0\ra$$
These q-bits are topologically protected from decoherence. In practice, a finite wire at $T>0$ can be realized as Kitaev’s Chain using a spin-orbit coupled wire with proximity-induced superconductivity and an external magnetic field. \cite{NadjPerge}.

Since the energy of Bogoliubons is never negative, see \cref{eqBVtrans}, the ground state $|Gs.\ra$ has no Bogoliubons, that is to say,
$$|Gs.\ra = |0\ra;\quad\textrm{where }\gamma_k|0\ra=0\,\forall k$$
The $n$\ts{th} excitation, $\gamma_{k_1}^\dagger\gamma_{k_2}^\dagger\dots\gamma_{k_n}^\dagger|0\ra$, corresponds to excitations in the quantum Ising model, viz., domain-wall formation in the ferromagnetic phase and spin-flip in the paramagnetic phase. The ground state in terms of $c_k$-fermions can be calculated by writing wave-function as an arbitrary combination of Cooper pairs
$$|Gs.\ra ={\cal N}\prod\limits_{q}e^{\alpha_qc_{-q}^\dagger c_{q}}|0\ra_{c}$$
Here we are using subscript $c$ to make a distinction between Bogoliubons vacuum state and $c$-fermions (fermions of Kitaev’s chain) vacuum state.
With property
$$\gamma_k|Gs.\ra=0 \implies u_kc_k |Gs.\ra = v_kc_{-k}^\dagger |Gs.\ra$$
we get,
$$|Gs.\ra = u_k^2\prod\limits_{k}(1 + \psi_{Cp.}(k) c_{-k}^\dagger c_{k})|0\ra_{c}$$
The term $\psi_{Cp.}(k)$ can be loosely interpreted as the wave-function of Cooper pairs. In real space,
$$|\psi_{Cp.}(x)|=\bigg|\int\limits_ke^{ipx}\psi_{Cp.}(k)\bigg|\sim\begin{cases}
                                                      e^{-|x|/\zeta}, & g\gg 1\\
                                                      \mathrm{const.}, & g\to 0
                                                     \end{cases}
$$
The limit $g\to 0$ corresponds to the weak-pairing of Cooper pairs of infinite size. The weak pairing is topologically non-trivial. Whereas, the limit of $g\gg 1$ corresponds to strong-pairing Cooper pairs over a length scale of $ \zeta $. The strong-pairing is topologically trivial.

\section{Conclusion}
This general article's primary focus is mapping a transverse (quantum) Ising model in one-dimensional to the Kitaev Chain to show how a thermodynamic phase transition problem, albeit at zero temperature, reduces to a problem of topological phase transition.

We end our discussion with a remark. Landau's theory has been and continues to be extremely useful in the usual paradigm of phase transitions, where an order-parameter characterizes phases. However, we see the difficulty of using Landau's theory in the Topological phase transition, where phases are characterized by different topological invariant \cite{Reso3}. The value of topological invariant changes discontinuously at the transition point, and they are not as useful as the order parameter. Nevertheless, the theory of critical phenomena like diverging correlation length and diverging correlation time is not different. We have restricted our discussion to strictly zero temperature. Nevertheless, the quantum phase transition leaves their signature at finite temperatures is now well known \cite{CONTINENTINO20141561}.

Other kinds of topological phase transitions referred to as Lifshitz transitions, where bands in a solid merge or disappear below the Fermi surface, as an external control parameter is varied \cite{Lifshitz}. The Landau approach is again of no use here.

\appendix
\section*{Appendix}
\section{Mapping transverse Ising model to classical Ising model}\label{mapquanttoclass}
In the following section, we will map one-dimensional transverse Ising model to two-dimensional classical Ising model by introducing imaginary time-slicing. 

\begin{figure}[h]
 \caption{Renormalization of a transverse-field Ising model in two dimensions.}
 \label{figure1}
 \centering
 \includegraphics[width=\columnwidth]{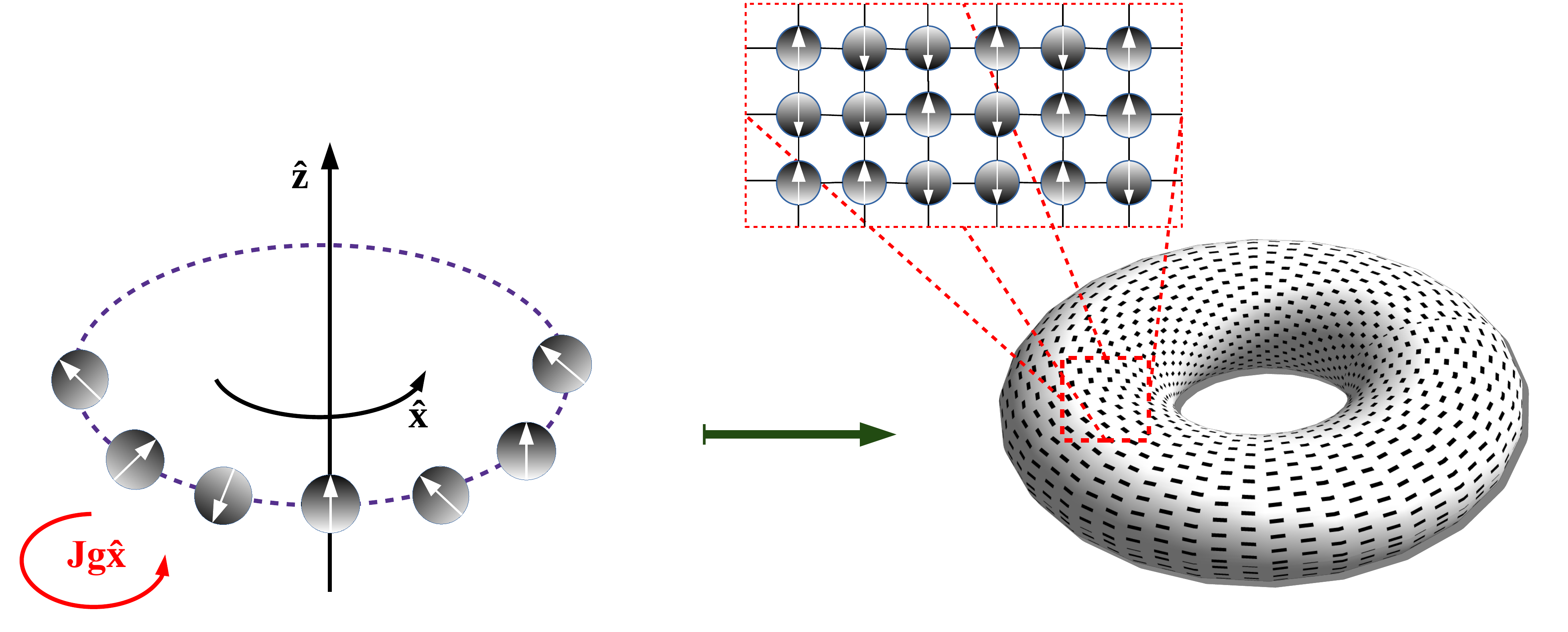}
\end{figure}

There are two terms in the transverse Ising Hamiltonian
$${\cal H} = {\cal H}_0 + {\cal H}_1$$
$${\cal H}_0=-J\sum\limits_{\la ij\ra}\hat{\sigma}_i^z\hat{\sigma}_j^z\quad \textrm{and}\quad {\cal H}_1=-Jg\sum\limits_{i}\hat{\sigma}_i^x$$

We can write the partition function ${\cal Z}$ by slicing the inverse temperature $\beta$ into several parts as
\begin{align*}
 {\cal Z}=&\textrm{Tr}e^{-\beta{\cal H}}\\
    =&\textrm{Tr}[e^{-\Delta\tau{\cal H}}e^{-\Delta\tau{\cal H}}\dots e^{-\Delta\tau{\cal H}}];\quad (\Delta\tau=\beta/T)\\
    =&\sum\limits_{\{S^z\}}\la S^z|e^{-\Delta\tau{\cal H}}e^{-\Delta\tau{\cal H}}\dots e^{-\Delta\tau{\cal H}}|S^z\ra
\end{align*}
where $|S^z\ra\equiv |s_1^z\ra\otimes |s_2^z\ra\otimes \dots|s_N^z\ra$, and $|s_i^z\ra$ is the spin state at lattice site $i$ on the Ising chain. We will insert the identity operator $\mathbb I=\sum\limits_{\{S^z\}}|S^z\ra\la S^z|$ between every exponential term $e^{-\Delta\tau{\cal H}}$ in partition function, so that,
$${\cal Z}=\sum\limits_{\{S_{t}^z\}}\la S_{1}^z|e^{-\Delta\tau{\cal H}}|S_{T}^z\ra\la S_{T}^z|e^{-\Delta\tau{\cal H}}|S_{T-1}^z\ra\dots\la S_{2}^z| e^{-\Delta\tau{\cal H}}|S_{1}^z\ra$$
The index $t$ in $|S_{t}^z\ra$ is referred as imaginary-time index. In the limit $\Delta\tau J\ll 1$, we can expand the exponential term $e^{-\Delta\tau{\cal H}_0-\Delta\tau{\cal H}_1}$ by using Suzuki-Trotter approximation as
$$e^{-\Delta\tau{\cal H}_0-\Delta\tau{\cal H}_1}=e^{-\Delta\tau{\cal H}_0}e^{-\Delta\tau{\cal H}_1}+{\cal O}(\Delta\tau^2[{\cal H}_0,{\cal H}_1])$$
Since, ${\cal H}_0$ acts on eigen state of Pauli matrices $\hat\sigma^z$, we can evaluate the term $\la S_{t+1}^z| e^{-\Delta\tau{\cal H}}|S_{t}^z\ra$ as
$$\la S_{t+1}^z| e^{-\Delta\tau{\cal H}}|S_{t}^z\ra=e^{-\Delta\tau J\sum\limits_{i=1}^{N}s_{i,t}^zs_{i+1,t}^z}\la S_{t+1}^z|e^{-\Delta\tau h\sum\limits_{i=1}^{N}\hat\sigma_i^x}|S_{t}^z\ra $$
Using the identity $e^{\Delta\tau h\hat\sigma_i^x}=\mathbb{I}\cosh(\Delta\tau h)+\hat\sigma_i^x\sinh(\Delta\tau h)$ we can write
\begin{align*}
\la \uparrow|e^{\Delta\tau h\sigma_i^x}|\uparrow\ra&=\cosh(\Delta\tau h)=\la \downarrow|e^{\Delta\tau h\sigma_i^x}|\downarrow\ra\equiv\Lambda e^\gamma\\
\la \uparrow|e^{\Delta\tau h\sigma_i^x}|\downarrow\ra&=\sinh(\Delta\tau h)=\la \downarrow|e^{\Delta\tau h\sigma_i^x}|\uparrow\ra\equiv\Lambda e^{-\gamma}
\end{align*}
where
$$\gamma=-\dfrac{1}{2}\log(\tanh(\Delta\tau h))\quad\mathrm{and}\quad\Lambda^2=\sinh(\Delta\tau h)\cosh(\Delta\tau h)$$
The term $\la S_{t+1}^z| e^{-\Delta\tau{\cal H}}|S_{t}^z\ra$ simplifies to
$$\la S_{t+1}^z| e^{-\Delta\tau{\cal H}}|S_{t}^z\ra=\Lambda^Ne^{\Delta\tau J\sum\limits_{i=1}^{N}s_{i,t}^zs_{i+1,t}^z+\gamma\sum\limits_{i=1}^{N}s_{i,t}^zs_{i,l+1}^z}$$
And the partition function simplifies to
$${\cal Z}=\Lambda^{NT}\sum\limits_{\{s_{i,t}^z\}}e^{\Delta\tau J\sum\limits_{i=1}^{N}\sum\limits_{t=1}^{T}s_{i,t}^zs_{i+1,t}^z+\gamma\sum\limits_{i=1}^{N}\sum\limits_{l=1}^{T}s_{i,t}^zs_{i,t+1}^z}$$
The partition function above can be identified as the partition function for a two-dimensional anisotropic classical Ising model with couplings $\beta J_x=\Delta\tau J$ and $\beta J_y=\gamma$. With this, we complete our discussion on Classical-Quantum correspondence of the
Ising model.

\begin{acknowledgments}
The author is thankful to Prof. Dr. PK Mohanty and Dr. Goutam Sheet for their guidance. The author would like to express his gratitude to an anonymous reviewer who has helped enrich this manuscript. The author is thankful to Abhishikta for her help in reducing grammatical errors. The author would also like to thank his parents and his sister for their support.
\end{acknowledgments}

\bibliographystyle{unsrt}
\bibliography{mybib}

\end{document}